\documentclass[twocolumn]{aastex63}
\usepackage{amsmath,latexsym}    
\usepackage{graphicx}   
\usepackage{verbatim}   
\usepackage{color}      
\usepackage{subfigure}  
\usepackage{hyperref}   
\usepackage{tensor}     
\usepackage{amssymb}
\usepackage{enumerate}	
\usepackage{braket}		
\usepackage{comment}

\newcommand{\mpcoh}{\,h^{-1}\,{\rm Mpc}}

\newcommand{\nn}{\nonumber}

\newcommand{\bfk}{\boldsymbol{k}}
\newcommand{\bfx}{\boldsymbol{x}}

\newcommand{\bfr}{\boldsymbol{r}}

\newcommand{\xiggs}{\xi_{gg}^s}
\newcommand{\xigps}{\xi_{g+}^s}
\newcommand{\xips}{\xi_{+}^s}
\newcommand{\xims}{\xi_{-}^s}

\newcommand{\xiXs}{\xi_{X}^s}

\newcommand{\PXs}{P_{X}^s}

\newcommand{\xiggls}{\xi_{gg,\ell}^s}
\newcommand{\xigpls}{\xi_{g+,\ell}^s}
\newcommand{\xipls}{\xi_{+,\ell}^s}
\newcommand{\ximls}{\xi_{-,\ell}^s}
\newcommand{\xipmls}{\xi_{\pm,\ell}^s}

\newcommand{\xiXls}{\xi_{X,\ell}^s}

\newcommand{\be}{\begin{equation}}
\newcommand{\ee}{\end{equation}}

\newcommand{\bftheta}{{\boldsymbol \theta}}

\graphicspath{{./}{figures/}}

\submitjournal{ApJ Letters}

\shortauthors{Okumura \& Taruya}

\begin{document}

\title{First Constraints on Growth Rate from Redshift-Space Ellipticity Correlations of SDSS Galaxies at $0.16 < z < 0.70$}
\email{tokumura@asiaa.sinica.edu.tw}

\author{Teppei Okumura}
\affiliation{Institute of Astronomy and Astrophysics, Academia Sinica, No. 1, Section 4, Roosevelt Road, Taipei 10617, Taiwan}
\affiliation{Kavli Institute for the Physics and Mathematics of the Universe (WPI), UTIAS, The University of Tokyo, Chiba 277-8583, Japan}

\author{Atsushi Taruya}
\affiliation{Center for Gravitational Physics and Quantum Information, Yukawa Institute for Theoretical Physics, Kyoto University, Kyoto 606-8502, Japan}
\affiliation{Kavli Institute for the Physics and Mathematics of the Universe (WPI), UTIAS, The University of Tokyo, Chiba 277-8583, Japan}

\begin{abstract}

We report the first constraints on the growth rate of the universe,
$f(z)\sigma_8(z)$, with intrinsic alignments (IAs) of galaxies.  We
measure the galaxy density-intrinsic ellipticity cross-correlation and
intrinsic ellipticity autocorrelation functions over $0.16 < z < 0.7$
from luminous red galaxies (LRGs) and LOWZ and CMASS galaxy samples in
the Sloan Digital Sky Survey (SDSS) and SDSS-III BOSS survey. We
detect clear anisotropic signals of IA due to redshift-space
distortions.  By combining measured IA statistics with the
conventional galaxy clustering statistics, we obtain tighter
constraints on the growth rate.  The improvement is particularly
prominent for the LRG, which is the brightest galaxy sample and known
to be strongly aligned with underlying dark matter distribution; using the 
measurements on scales above $10\mpcoh$, we
obtain $f\sigma_8 = 0.5196^{ + 0.0352}_{ - 0.0354}$ (68\% confidence level) from
the clustering-only analysis and $f\sigma_8 = 0.5322^{ + 0.0293}_{ -
  0.0291}$ with clustering and IA, meaning $19\%$ improvement. The
constraint is in good agreement with the prediction of general
relativity, $f\sigma_8 = 0.4937 $ at $z=0.34$.  For LOWZ and CMASS
samples, the improvement of constraints on $f\sigma_8$ is found to be
$10\%$ and $3.5\%$, respectively.  Our results indicate that the contribution from IA statistics for cosmological constraints can be 
further enhanced by carefully selecting galaxies for a shape sample.
\end{abstract}





\section{Introduction} \label{sec:introduction}

Cosmological parameters have been precisely determined via various
observations: cosmic microwave background
\citep{Planck-Collaboration:2020}, large-scale structure of the
universe \citep{Alam:2017}, and gravitational lensing
\citep{Hikage:2019}. However, the origin of the accelerating expansion
of the universe, namely, dark energy or/and modification of Einstein's
gravity theory, is still a complete mystery in fundamental physics.
Thus, deeper and wider galaxy surveys are ongoing to better understand
the expansion and growth history of the universe
\citep{Takada:2014,DESI-Collaboration:2016}.

In parallel, we need to keep exploring methods that maximize the use
of cosmological information encoded in given observations.  There is a
growing interest in using intrinsic alignment (IA) of galaxy shapes
\citep{Croft:2000,Heavens:2000,Hirata:2004} as a
geometric and dynamical probe of cosmology complimentary to galaxy
clustering.  Although there are various observational studies of IA,
they mainly focused on the contamination to weak gravitational-lensing
measurements \citep[e.g.,][]{Mandelbaum:2006,Okumura:2009,Joachimi:2011,Li:2013,Singh:2015, Tonegawa:2022}. The
anisotropy of three-dimensional IA statistics has been detected by
\cite{Singh:2016}. The full cosmological information of IA, however,
had not been investigated at that time.

To fully exploit cosmological information encoded in anisotropic IA,
theoretical modeling of the three-dimensional IA correlations has been
developed \citep{Okumura:2020,Okumura:2020a,Kurita:2021}.
A series of our papers
\citep{Taruya:2020,Chuang:2022,Okumura:2022} has also shown that the
three-dimensional IA statistics in redshift space provide additional
constraints on the linear growth rate of the universe,
$f=d\ln{\delta_m}/d\ln{a}$ ($a$ and $\delta_m$ being the scale factor
and matter density perturbation), which is used to test modified
gravity models.
Furthermore, recent studies showed that IA can be used as probes of not only modified gravity models but also other effects such as 
primordial non-Gaussianity, neutrino masses, and gravitational redshifts \citep{Schmidt:2015,Lee:2023,Zwetsloot:2022,Saga:2023}.

In this paper, besides conventional galaxy density correlation
functions, we measure intrinsic ellipticity correlation functions from
various galaxy samples in the Sloan Digital Sky Survey (SDSS) and
SDSS-III Baryon Oscillation Spectroscopic Survey (BOSS).  We then
present the first joint constraints on the growth rate from the galaxy
IA and clustering.  Otherwise stated, we assume a flat $\Lambda$CDM
model determined by \cite{Planck-Collaboration:2020} as our fiducial
cosmology throughout this paper.


\section{SDSS Galaxy Samples}\label{sec:sdss}

We analyze the galaxy distribution over $0.16\leq z \leq 0.70$ from
the SDSS-II \citep{Eisenstein:2001} and SDSS-III BOSS
\citep{Reid:2016}.  First, we use the luminous red galaxy (LRG) sample ($0.16 \leq z \leq
0.47$) from the SDSS Data Release 7 (DR7).  Galaxies in the sample
have rest-frame $g$-band absolute magnitudes, $-23.2<M_g<-21.2$
($H_0=100\,{\rm km}~{\rm s}^{-1}~{\rm Mpc}^{-1}$) with $K+E$
corrections of passively evolved galaxies to a fiducial redshift of
0.3.  The components of the ellipticity are defined as
\begin{align}
\left(
\begin{array}{c} \gamma_+ \\ \gamma_\times \end{array}
\right)(\bfx)
 = \frac{1-q^2}{1+q^2} 
\left(
\begin{array}{c} \cos{(2\beta_x)} \\ \sin{(2\beta_x)}\ \end{array}
\right)
\label{eq:ellip}
\end{align}
where $q$ is the minor-to-major-axis ratio ($0\leq q \leq 1$) and
$\beta_x$ is the position angle of the ellipticity from the north
celestial pole to east.  We use the ellipticity of LRG defined by the
$25 ~{\rm mag}~{\rm arcsec}^{-2}$ isophote in the $r$ band.  This LRG
sample is similar to that used in \citet{Okumura:2009} and
\cite{Okumura:2009a} but slightly extended from DR6 to DR7, 
with the total number of the LRG used being $105,334$.

We also use $353,804$ LOWZ ($0.16 \leq z \leq 0.43$) and $761,567$
CMASS ($0.43 \leq z \leq 0.70$) galaxy samples from the BOSS DR12.
For these samples, we adopt the ellipticity defined by
the adaptive moment \citep{Bernstein:2002}. While this method optimally corrects for the point-spread function (PSF) in the determined ellipticity, it is found to result in a small bias \citep{Hirata:2003}. The residual PSF remains in the shape
autocorrelation function at large scales \citep{Singh:2016}.  As we
show below, the correlation functions of these samples are very noisy,
and they do not contribute to cosmological constraints below.

As in our earlier studies, we set the axis ratio in
Equation~(\ref{eq:ellip}) to $q=0$
\citep{Okumura:2009a,Okumura:2009,Okumura:2019,Okumura:2020a}. We are
not interested in the amplitude of IA and marginalize it over. This
simplification will not affect results below.

\begin{figure*}[htbp]
\begin{center}
\includegraphics[width = 0.99\textwidth]{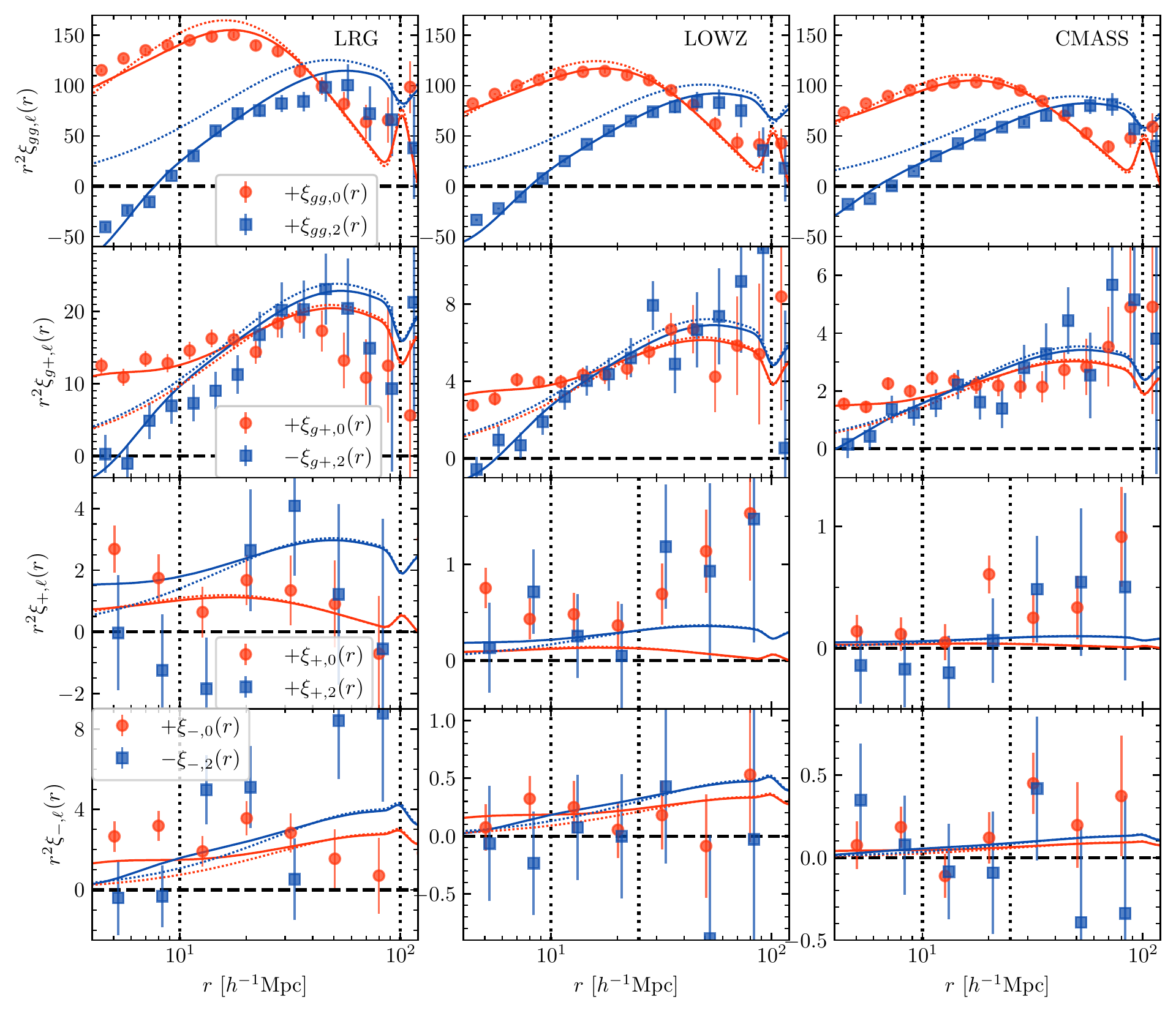}
\caption{Monopole and quadrupole correlation functions of SDSS
  galaxies in redshift space, $\xiggls$, $\xigpls$, $\xipls$, and
  $\ximls$ from top to bottom panels. The results for LRG, LOWZ, and
  CMASS samples are shown from the left to right panels, respectively.
  The error bars are estimated from jackknife resampling. The solid
  curves are the best-fit nonlinear alignment and RSD models jointly
  fitted for these four statistics, where the data points enclosed by
  the vertical lines are used. The dotted curves are the linear
  predictions as references. }
\label{fig:ximu_gg_gi_ii}
\end{center}
\end{figure*}


\section{Measurement of correlation functions}\label{sec:measurement}

In this section, we measure the redshift-space correlation functions
of galaxy density and IA from the SDSS samples, and estimate their
covariance matrix.

As a conventional clustering analysis, we use a galaxy
autocorrelation (GG) function in redshift space,
$\xiggs(\bfr)=\left\langle
\delta_g^s(\bfx_1)\delta_g^s(\bfx_2)\right\rangle$, where superscript
$s$ denotes the quantity defined in redshift space, $\bfr =
\bfx_2-\bfx_1$, and $\delta_g^s(\bfx)$ is the galaxy number density
fluctuation.  We adopt the \cite{Landy:1993} estimator to measure it,
\be
\xiggs(\bfr) = \frac{(D-R)^2}{RR} = \frac{DD-2DR+RR}{RR}, \label{eq:ls}
\ee
where $DD$, $RR$, and $DR$ are the normalized counts of galaxy--galaxy,
random--random, and galaxy--random pairs, respectively.  We then obtain
the multipole moments,
\be
\xiggls(r) = (2\ell + 1) \int^1_0 d\mu ~\xiggs(r,\mu_{\bfr}){\cal L}_\ell (\mu_{\bfr}) , \label{eq:multipole}
\ee
where $r = |\bfr |$, $\mu_{\bfr}$ is the direction cosine between the
line of sight and $\bfr$, and ${\cal L}_\ell$ is the $\ell$th-order
Legendre polynomials.
To obtain the multipoles via Eq.~(\ref{eq:multipole}), we estimate $\xiggs(r,\mu_{\bfr})$ 
with the angular bin size of $\Delta\mu_{\bfr}=0.1$ in Eq.~(\ref{eq:ls}) and take the sum over $\mu_{\bfr}$.

The first row of Figure~\ref{fig:ximu_gg_gi_ii} shows multipole
moments of the GG correlation functions.  The first, second, and third
columns show the results from the LRG, LOWZ, and CMASS samples,
respectively.  Since the hexadecapole is noisy, we analyze only the
monopole and quadrupole moments \citep{Kaiser:1987}. These correlation
functions have been measured in various previous works
\citep[e.g.,][]{Samushia:2012,Alam:2017} and our measurements are consistent
with theirs.

Next, we introduce intrinsic alignment statistics, which are
density-weighted quantities. The galaxy position--intrinsic
ellipticity (GI) correlation, $\xigps$, and intrinsic
ellipticity--ellipticity (II) correlations, $\xips$ and $\xims$, are
defined by
\be
\xiXs(\bfr) = \left\langle \left[1+\delta_g(\bfx_1)\right] \left[1+\delta_g(\bfx_2)\right] W_X(\bfx_1,\bfx_2)\right\rangle,
\ee
where $W_{g+}(\bfx_1,\bfx_2) = \gamma_+(\bfx_2)$ and
$W_{\pm}(\bfx_1,\bfx_2) = \gamma_+(\bfx_1)\gamma_+(\bfx_2) \pm
\gamma_\times(\bfx_1)\gamma_\times(\bfx_2)$.  For the II correlations,
we label $\xips$ and $\xims$ individually as II$(+)$ and II($-$)
correlations, respectively.  The GI correlation function is estimated
as \citep{Mandelbaum:2006},
\begin{align}
&\xi_{g+}(\bfr) = \frac{S_+(D-R)}{RR} = \frac{S_+D-S_+R}{RR} \ , 
\end{align}
where $S_+D$ is the sum over all pairs with separation $\bfr$ of the
$+$ component of the ellipticity, $S_+D = \sum_{i\neq j| \bfr}
{\gamma_+(j|i)}$, with $\gamma_+(j|i)$ being the ellipticity of galaxy
$j$ measured relative to the direction to galaxy $i$, and $S_+R$ is
defined similarly.  The II correlation functions are estimated as
\begin{align}
&\xi_{\pm}(\bfr) 
=\frac{S_+S_+ \pm S_\times S_\times}{RR}\ , 
\end{align}
where $S_+S_+ = \sum_{i\neq j| \bfr} {\gamma_+(j|i)\gamma_+(i|j)}$ and
similarly for $S_\times S_\times$.  Finally, multipole moments for the
IA correlations, $\xigpls$ and $\xipmls$, are obtained via the same
equation as Equation~(\ref{eq:multipole}).  Again, since the
hexadecapole is noisy, we analyze only the $\ell=0$ and $\ell=2$
moments.

The second, third, and bottom rows of Figure~\ref{fig:ximu_gg_gi_ii}
respectively present redshift-space multipole moments of the GI,
II($+$) and II($-$) correlation functions.  Both the monopole and
quadrupole of the GI correlation are clearly detected in all the three
samples.  Particularly, LRG are the brightest galaxy sample and shows
the strongest signal because IA has a strong luminosity dependence.
Though LOWZ has a redshift range similar with LRG, it targets fainter
galaxies and thus has higher number density.  Therefore, the LOWZ
sample shows lower GI amplitude, confirming the earlier detection by
\cite{Singh:2016}. We find even a lower GI signal in the CMASS sample.
The monopole of the II correlation is clearly detected for the LRG
sample, as in \cite{Okumura:2009}, while the newly measured quadrupole
is noisier and consistent with zero.  Those for the LOWZ and CMASS
samples have much lower amplitude, and are somewhat consistent with
zero.  Furthermore, their shapes are determined by the adaptive moment
and have nonzero correlation due to the PSF at $r > 30\mpcoh$
\citep{Singh:2016}.

We estimate the covariance matrix for the measured correlation
functions, ${\rm C}_{ij}^{X_\ell X'_{\ell'}}\equiv {\rm C}\left[
  \xi_{X,\ell}(r_i), \xi_{X',\ell'}(r_j) \right]$, with $X=\{ gg, g+,
+, - \}$ and $\ell=\{0,2\}$, using the jackknife resampling method.
While jackknife is not an unbiased error estimator, it provides
reliable error bars for the statistics whose error is dominated by the
shape noise \citep{Mandelbaum:2006}.  The error bars shown in
Figure~\ref{fig:ximu_gg_gi_ii} are the square root of the diagonal
components of the covariance matrix.


\section{Theoretical prediction}
\label{sec:theory}

Here we present theoretical models to interpret the measured
correlation functions.  Since theoretical models are naturally
provided in Fourier space, we first present models for the power
spectra, $\PXs$, perform the Fourier transform,
\be
\xiXs(\bfr) = \int \frac{d^3 \bfk}{(2\pi)^3}\PXs(\bfk)e^{i\bfk\cdot\bfr},
\ee
where $X = \{gg,g+,+,-\}$, and obtain the multipole moments $\xiXls$
via Equation~(\ref{eq:multipole}).

\subsection{Galaxy correlations}

For the galaxy power spectrum, we adopt nonlinear redshift-space
distortion (RSD) model proposed by
\citep{Scoccimarro:2004,Taruya:2010},
\begin{align}
P_{gg}^s(\bfk) = \left[ b^2 P_{\delta\delta}(k)   + 2bf\mu_{\bfk}^2 P_{\delta\Theta}  \right.  
&  (k)
\nn \\  \left.
+ f^2\mu_{\bfk}^4 P_{\Theta\Theta}(k) \right] 
& D_{\rm FoG}^2(k\mu_{\bfk}\sigma_v), \label{eq:scoccimarro}
\end{align}
where $k=|\bfk|$, $\mu_{\bfk}$ is the direction cosine between the
observer's line of sight and the wavevector $\bfk$, and $b$ the galaxy
bias.  The quantities $P_{\delta\delta}$ and $P_{\Theta\Theta}$ are
the nonlinear autopower spectrum of density and velocity fields,
respectively, and $P_{\delta\Theta}$ is the their cross-power
spectrum.  We adopt the revised \texttt{Halofit} model to compute
$P_{\delta\delta}$ \citep{Takahashi:2012}, and then $P_{\delta\Theta}$
and $P_{\Theta\Theta}$ are computed using the fitting formulae derived
by \cite{Hahn:2015}.  The function $D_{\rm FoG}$ is a damping function
due to the Finger-of-God (FoG) effect characterized by the nonlinear
velocity dispersion parameter $\sigma_v$.  We adopt a simple Gaussian
function, $D_{\rm
  FoG}(k\mu_{\bfk}\sigma_v)=\exp{\left(-k^2\mu_{\bfk}^2\sigma_v^2/2\right)}$.
With this Gaussian function, the nonlinear multipoles are expressed
analytically by a simple Hankel transform \citep{Taruya:2009}. In the
linear-theory limit, $P_{\delta\delta} = P_{\delta\Theta} =
P_{\Theta\Theta}$ and $D_{\rm FoG}=1$, and hence
Equation~(\ref{eq:scoccimarro}) converges to the original Kaiser
formula.  Since $P_{\delta\delta}$, $P_{\delta\Theta}$ and
$P_{\Theta\Theta}$ are proportional to the square of the normalization
parameter of the density fluctuation, $\sigma_8^2(z)$, free parameters
for this model are $\bftheta=(b\sigma_8, f\sigma_8,\sigma_v)$.

\subsection{Intrinsic alignment correlations}

To quantify the cosmological information encoded in the IA statistics,
we consider the LA model, which assumes a linear relation between the
intrinsic ellipticity and tidal field \citep{Catelan:2001}. In Fourier
space, the ellipticity projected along the line of sight ($z$-axis) is
given by
\begin{align}
\left(
\begin{array}{c} \gamma_+ \\ \gamma_\times \end{array}
\right) (\bfk)=b_K
\left(
\begin{array}{c} 
    (k_{x}^{2}-k_{y}^{2})/k^2 \\
    2k_{x}k_{y}/k^2 
\end{array}
\right)
\delta_m(\bfk),
\end{align}
where $b_{K}$ represents the redshift-dependent coefficient of the
intrinsic alignments, which we refer to as the shape bias.
We adopt the nonlinear alignment (NLA) model,
which replaces the linear matter density field $\delta_m$ by the
nonlinear one \citep{Bridle:2007}. Furthermore, the redshift-space
shape field is multiplied by the damping function due to the FoG
effect.

Adopting also the nonlinear RSD model in Equation
(\ref{eq:scoccimarro}), the GI and II power spectra are expressed as
\begin{align}
P_{g+}^s(\bfk) &= b_Kk^{-2}(k_x^2 - k_y^2)\left\{b P_{\delta\delta}(k) +f\mu_{\bfk}^2P_{\delta\theta}(k)\right\} \nn \\
&\qquad \qquad \qquad \qquad \qquad \times D_{\rm FoG}^2(k\mu_{\bfk} \sigma_v),\label{eq:nla_rsd_gi}  \\
P_{\pm}^s(\bfk) &= b_K^2 k^{-4}\left[ (k_x^2 - k_y^2)^2 \pm ( 2 k_x k_y)^2\right] P_{\delta\delta}(k) \nn \\
&\qquad \qquad \qquad \qquad \qquad \times D_{\rm FoG}^2(k\mu_{\bfk} \sigma_v)~. \label{eq:nla_rsd_ii}
\end{align}
Note that \cite{Singh:2015} showed that the shape field is insensitive to RSD.
While it is true in the linear RSD model,
the FoG effect comes into IA power spectra in the same way as the GG spectrum because it is caused purely by a coordinate transform from real to redshift space (T.~Okumura et al. 2023, in preparation). 

Similarly to $\xiggls$, multipole moments of the IA correlations,
$\xigpls$ and $\xipmls$, can be expressed by a Hankel transform.  Since correlation
functions of the projected shape are naturally expressed by the
associated Legendre polynomial basis \citep{Kurita:2022}, the
nonlinear model of $\xigpls$ and $\ximls$ involving the FoG factor produces infinite series for
each Legendre multipole.  We computed the expansion up to the 12th
order and confirmed the convergence of the formula.  
The nonlinear model of $\xipls$ has a form similar with $\xiggls$. 
We have four free parameters for the IA statistics,
${\bftheta}=(b\sigma_8,b_K\sigma_8, f\sigma_8,\sigma_v)$. 
Taking the
linear-theory limit of the GI and II correlation functions, namely $\sigma_v \to 0$
limit in Equations~(\ref{eq:nla_rsd_gi}) and (\ref{eq:nla_rsd_ii}), leads to the formulas presented
in \cite{Okumura:2020}. 
We will present the full
expressions of IA statistics with the Gaussian damping factor in our
upcoming paper.  


\section{Constraints on growth rate}\label{sec:result}

We perform the likelihood analysis and constrain the growth rate
parameter $f\sigma_8$ from the three SDSS galaxy
samples. Particularly, we show how well the constraints are improved
by combining IA statistics with the conventional galaxy clustering
statistics.  We compare the measured statistics, $\xiXls$, where
$X=\{gg,g+,+,-\}$ and $\ell=\{0,2\}$, to the corresponding
predictions.  The $\chi^2$ statistic is given by
\be
\chi^2(\bftheta) = \sum_{i,j,\ell,\ell',X,X'}{\Delta_i^{X_\ell} \left({\bf C}^{-1} \right)_{ij}^{X_\ell X'_{\ell '}} \Delta_j^{X'_{\ell'}} },
\ee
where $\Delta_i^{X_\ell} = \xi_{X,\ell}^{s,{\rm
    obs}}(r_i)-\xi_{X,\ell}^{s,{\rm th}}(r_i;\bftheta)$ is the
difference between the observed correlation function and theoretical
prediction with $\bftheta$ being a parameter set to be constrained.
The analysis is performed over the scales adopted, $r_{\rm min} \leq
r_i \leq r_{\rm max}$.  Since the jackknife method underestimates the
covariance at large scales, we set the maximum separation $r_{\rm
  max}=100\mpcoh$.  Moreover, as described in Sec.~\ref{sec:sdss}, the
II correlation functions of LOWZ and CMASS are affected by the
residual PSF at $r > 30\mpcoh$ \citep{Singh:2016}. We thus set $r_{\rm
  max}=25\mpcoh$ for the II correlations of these samples.  In
Appendix \ref{sec:scale_dependence}, we investigate how our
constraints change with $r_{\rm min}$, and we adopt $r_{\rm
  min}=10\mpcoh$.  
In Appendix \ref{sec:psf}, we provide further argument that 
our cosmological constraints are not biased by the effect of the uncorrected PSF.
For the clustering-only analysis, the covariance is
a $20\times20$ matrix, while for the full analysis of clustering and
IA, it is a $60\times60$ matrix for LRG and $48\times48$ for LOWZ and
CMASS.  The data points used for the analysis are enclosed by the
vertical lines in Figure~\ref{fig:ximu_gg_gi_ii}.

\begin{figure}[t]
  \includegraphics[width = 0.475\textwidth]{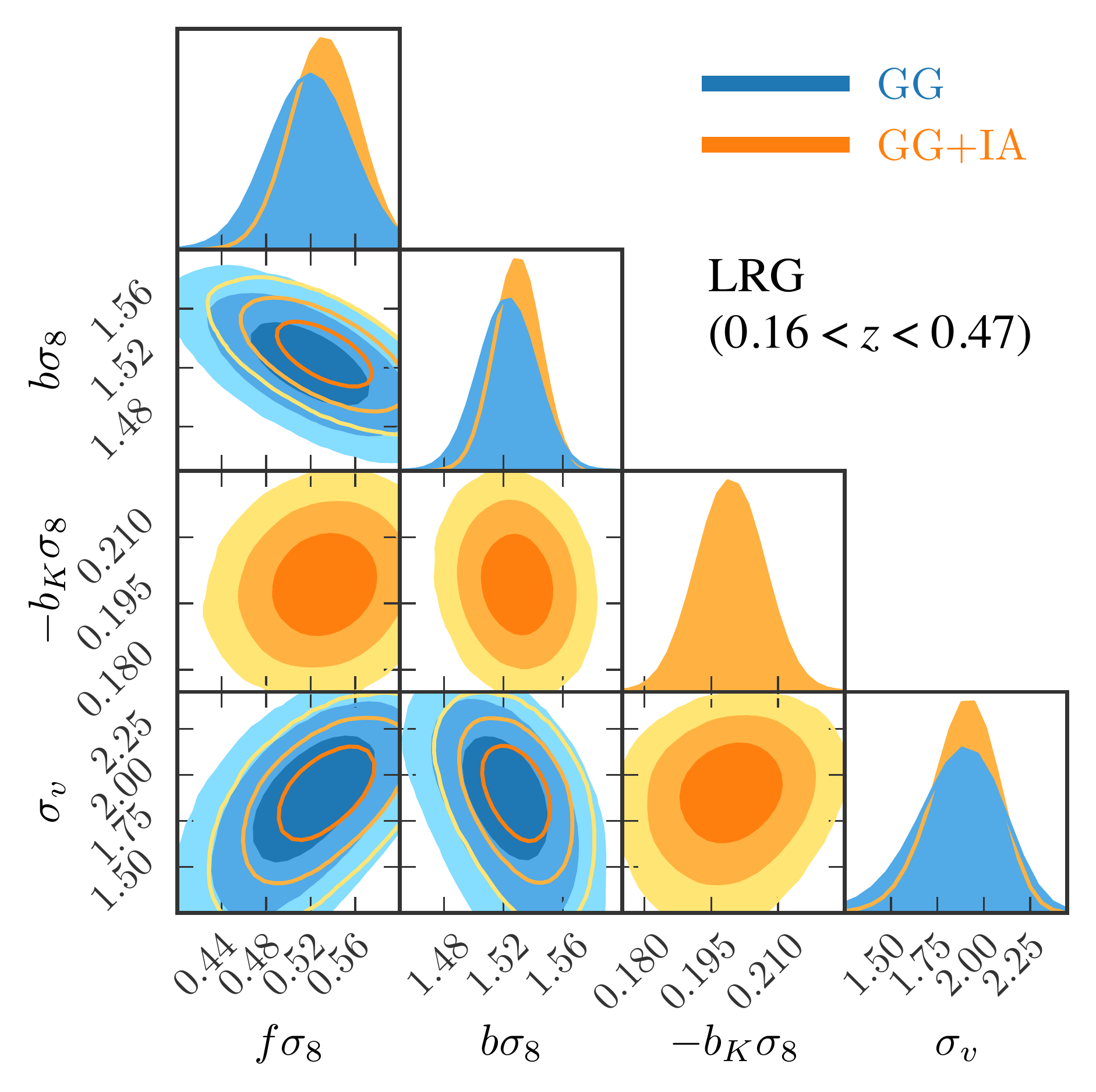}
\caption{Constraints on $(f\sigma_8,b\sigma_8,b_K\sigma_8,\sigma_v)$
  obtained from clustering-only analysis and combined analysis of
  clustering and IA, determined by the correlation functions of LRG
  sample at $10\leq r \leq 100\mpcoh$.  The contours show the $68\%,
  95\%$, and $99\%$ C.~L. from inward.  }
\label{fig:likelihood_lrg}
\end{figure}

\begin{figure*}[htbp]
\begin{center}
\includegraphics[width = 0.49\textwidth]{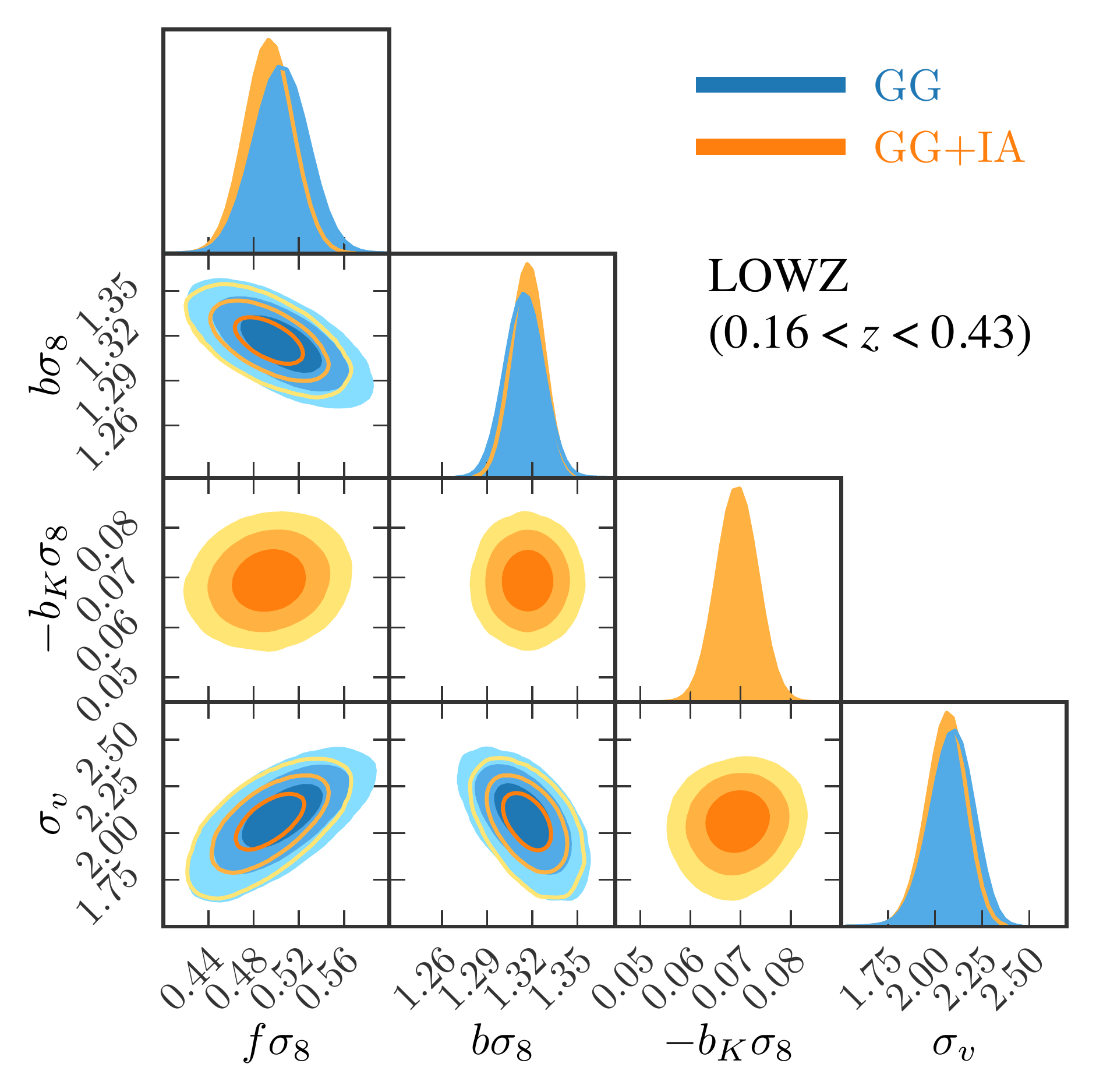}
\includegraphics[width = 0.49\textwidth]{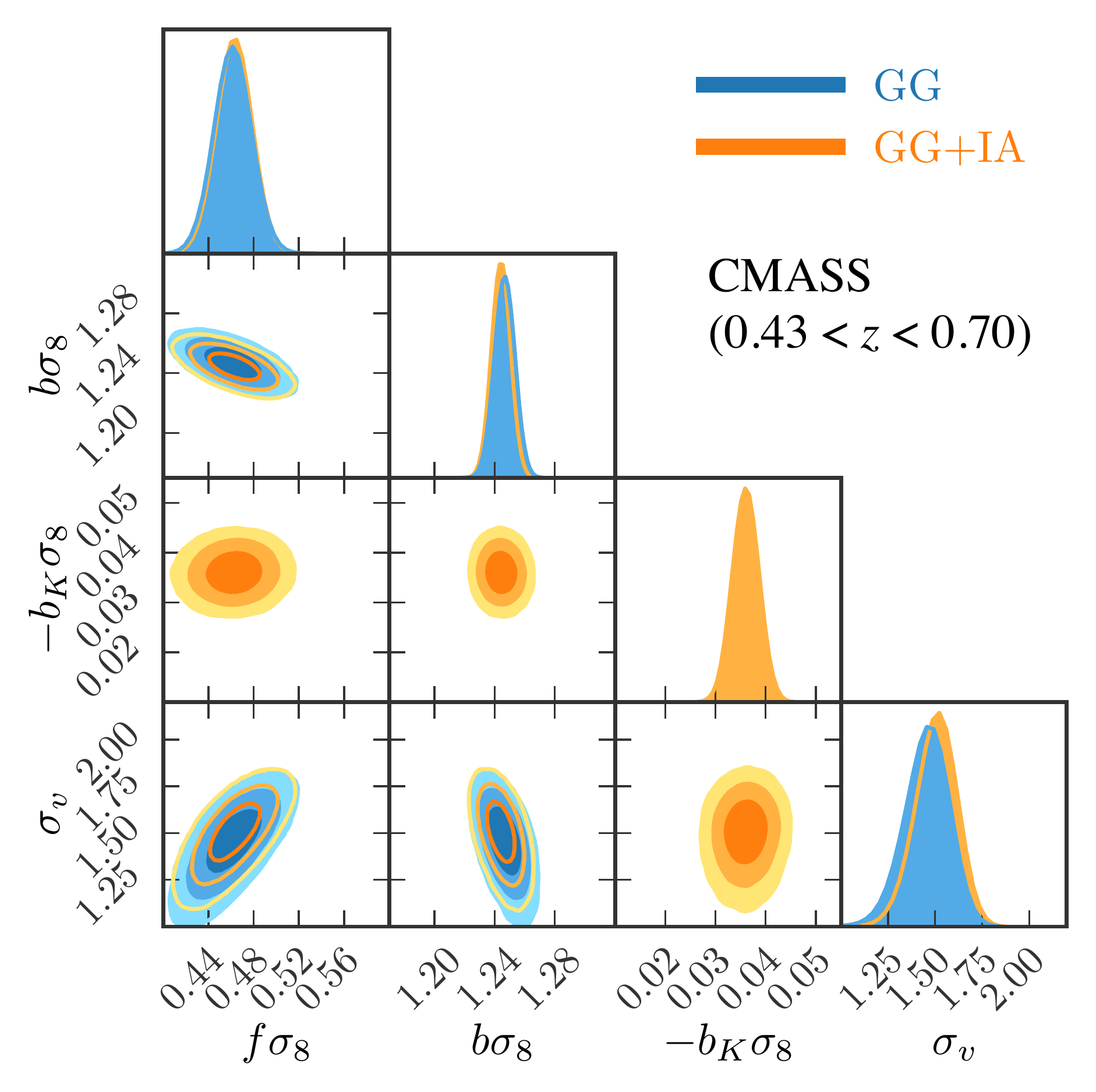}
\caption{Same as Figure~\ref{fig:likelihood_lrg}, but for LOWZ (top)
  and CMASS (bottom) samples. For these samples, the II correlations
  only at $10\leq r \leq 25\mpcoh$ are used, and the GG and GI
  correlations at $10\leq r \leq 100\mpcoh$ are used.}
\label{fig:likelihood_boss}
\end{center}
\end{figure*}

Figure~\ref{fig:likelihood_lrg} shows the parameter constraints
obtained from the LRG sample.  The blue and orange contours are
results with the clustering-only analysis and its combination with IA
statistics, respectively.  For the clustering-only analysis, after
marginalizing over $b\sigma_8$ and $\sigma_v$, we obtain the
constraint as $f\sigma_8 = 0.5196^{ + 0.0352}_{ - 0.0354}$ ($68\%$
confidence level).  For the combined analysis of clustering and IA, we obtain
$f\sigma_8 = 0.5322^{ + 0.0293}_{ - 0.0291}$ by further marginalizing
over the shape bias parameter $b_K$. Namely, the constraint on $f\sigma_8$ is
improved by $19\%$ by adding the IA statistics. Note that, as we set
$q=0$ in Equation~(\ref{eq:ellip}), the definition of $b_K$ here is
different from literature and one cannot directly compare the values.

\begin{figure}[tb]
\includegraphics[width = 0.47\textwidth]{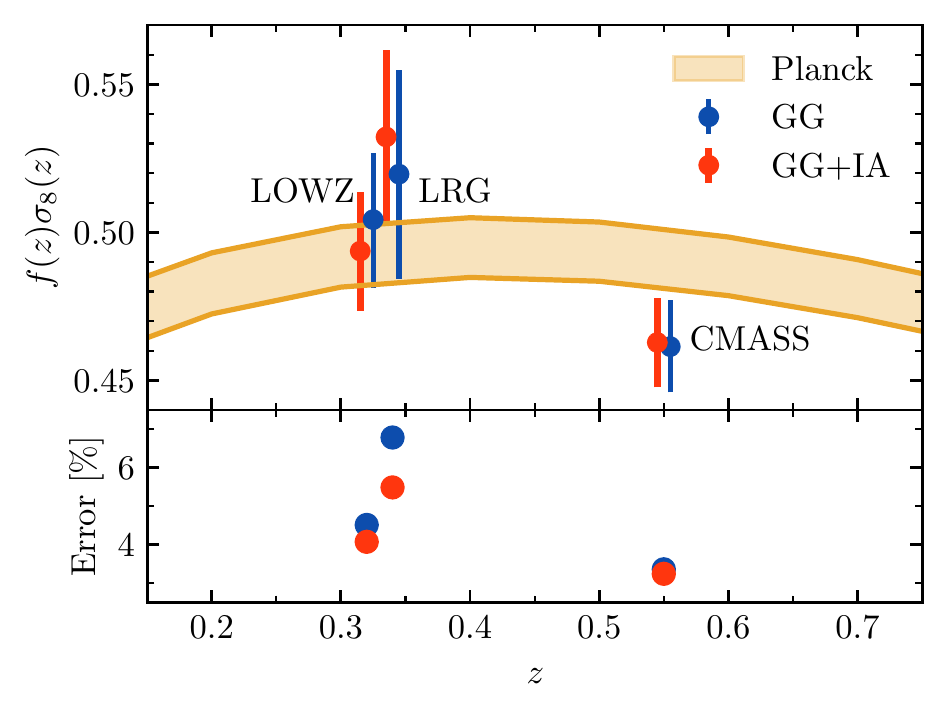}
\caption{Upper panel: Constraints on growth rate $f(z)\sigma_8(z)$
  from three SDSS galaxy samples compared to the best-fitting
  $\Lambda$CDM model from the Planck experiment. We adopt $r_{\rm
    min}=10\mpcoh$. Lower panel: $1\sigma$ error of the growth rate
  constraints, $\Delta (f\sigma_8) / f\sigma_8$. }
\label{fig:fsigma8z}
\end{figure}

The left and right panels of Figure~\ref{fig:likelihood_boss} show
results similar to Figure~\ref{fig:likelihood_lrg} but for LOWZ
and CMASS, respectively.  Using LOWZ, we obtain $f\sigma_8 = 0.5043^{+
  0.0226}_{ -0.0229}$ (GG only), and $f\sigma_8= 0.4937^{+ 0.0201}_{
  -0.0201}$ (GG$+$IA).  The LOWZ is a denser sample than the LRG by
targeting fainter galaxies, and thus, even the galaxy clustering alone
puts tighter constraints.  However, combining the IA statistics, LRG
provides almost as a strong constraint as LOWZ.  CMASS is also a
fainter population at higher redshift, $0.43 < z < 0.70$.  With the
GG-only analysis, we obtain $f\sigma_8 = 0.4614^{+ 0.0156}_{-
  0.0154}$, and with the GG+IA analysis, $f\sigma_8 = 0.4628^{+
  0.0149}_{- 0.0151}$.  Our analysis of these three galaxy samples
demonstrates that the contribution of IA to cosmological constraints
can be enhanced by adopting an optimal weighting to brighter galaxies
\citep{Seljak:2009a}. Exploring such an optimization is our future
work.

The best-fit nonlinear models jointly fitted for the clustering and IA statistics are shown by the solid curves in Figure~\ref{fig:ximu_gg_gi_ii}. Reduced $\chi^2$ values obtained for LRG, LOWZ, and CMASS samples are $\chi^2/\nu = 1.85, 1.14$, and $2.42$, respectively, where $\nu$ is the degree of freedom, $\nu=56$ for LRG and $\nu=44$ for LOWZ and CMASS.
The large $\chi^2$ value for the CMASS sample is due to small error bars in the GG correlation. If we adopt $r_{\rm min}=15\mpcoh$, the minimum $\chi^2$ is reduced to $\chi^2/\nu=1.68$. Accordingly, the best-fitting value of $f\sigma_8$ is shifted (see  Figure~\ref{fig:scale_dependence} in Appendix \ref{sec:scale_dependence}).

Finally, Figure~\ref{fig:fsigma8z} summarizes the constraints on
$f\sigma_8$ from the three galaxy samples we considered.  As shown in
the lower panel, the constraint gets tighter by adding IA statistics
to the galaxy clustering statistics.  Overall, the derived results are
consistent with the prediction of $\Lambda$CDM determined from the
Planck satellite experiment \citep{Planck-Collaboration:2020}.  
It indicates that combining IA and clustering
statistics enables us to obtain robust and tight constraints.


\section{conclusions} \label{sec:conclusion}

We have presented the first cosmological constraints using IA of the
SDSS galaxies.  We have measured the redshift-space GI and II
correlation functions of LRG, LOWZ, and CMASS galaxy samples.  By
comparing them with the models of nonlinear alignment and RSD effects,
we have constrained the growth rate of the density perturbation,
$f(z)\sigma_8(z)$. We found that combining IA with clustering enhances
the growth rate constraint by $\sim 19\%$ compared to the
clustering-only analysis for the LRG sample.  This improved constraint
on $f\sigma_8$ is only slightly worse than
that obtained from the LOWZ, which is a
much denser sample by targeting fainter galaxies. This indicates a
potential that the contribution of the IA statistics can be further
enhanced by adopting an optimal weighting to brighter galaxies.

In this work we considered only the dynamical constraint via RSD.
However, baryon acoustic oscillations (BAOs) observed in the galaxy
distributions \citep{Eisenstein:2005} were shown to be
also encoded in galaxy IA statistics and thus useful to tighten
geometric constraints \citep{Chisari:2013,Okumura:2019}. The
cosmological analysis of IA simultaneously using RSD and BAO will be
shown in our future work.

The benefits of using IA can be further enhanced by improving the
model.  In this paper we worked with a simple extension of the NLA
model to include partly the FoG effect (T.~Okumura et al. 2023, in preparation). However, more sophisticated
nonlinear models of IA statistics have been proposed recently
\citep{Blazek:2019,Vlah:2020,Akitsu:2021,Matsubara:2022}.  These
models enable us to use the measured IA correlation functions down to
smaller scales, which will enhance the science return from IA of
galaxies.

\acknowledgments  T.O. thanks Ting-Wen Lan and Hironao Miyatake for useful discussion on how to treat photometric information from the SDSS server. We also thank the referee for the careful reading and suggestions. We are grateful for the Yukawa Institute for Theoretical Physics at Kyoto University for discussions during the YITP workshop YITP-W-22-16 on ``New Frontiers in Cosmology with the Intrinsic Alignments of Galaxies,'' which was useful to complete this work. 
T.O. acknowledges support from the Ministry of Science and Technology of Taiwan under grants Nos. MOST 110-2112-M-001-045- and 111-2112-M-001-061- and the Career Development Award, Academia Sinica (AS-CDA-108-M02) for the period of 2019-2023. A.T. acknowledges the support from MEXT/JSPS KAKENHI grant No. JP20H05861 and JP21H01081, and Japan Science and Technology Agency AIP Acceleration Research grant No. JP20317829. Numerical computations were carried out partly at Yukawa Institute Computer Facility. 
Funding for SDSS-III has been provided by the Alfred P. Sloan Foundation, the Participating Institutions, the National Science Foundation, and the U.S. Department of Energy Office of Science. The SDSS-III website is http://www.sdss3.org/.

\appendix 


\section{Scale dependence of parameter constraints}\label{sec:scale_dependence}

\begin{figure*}[bt] \begin{center} \includegraphics[width = 0.99\textwidth]{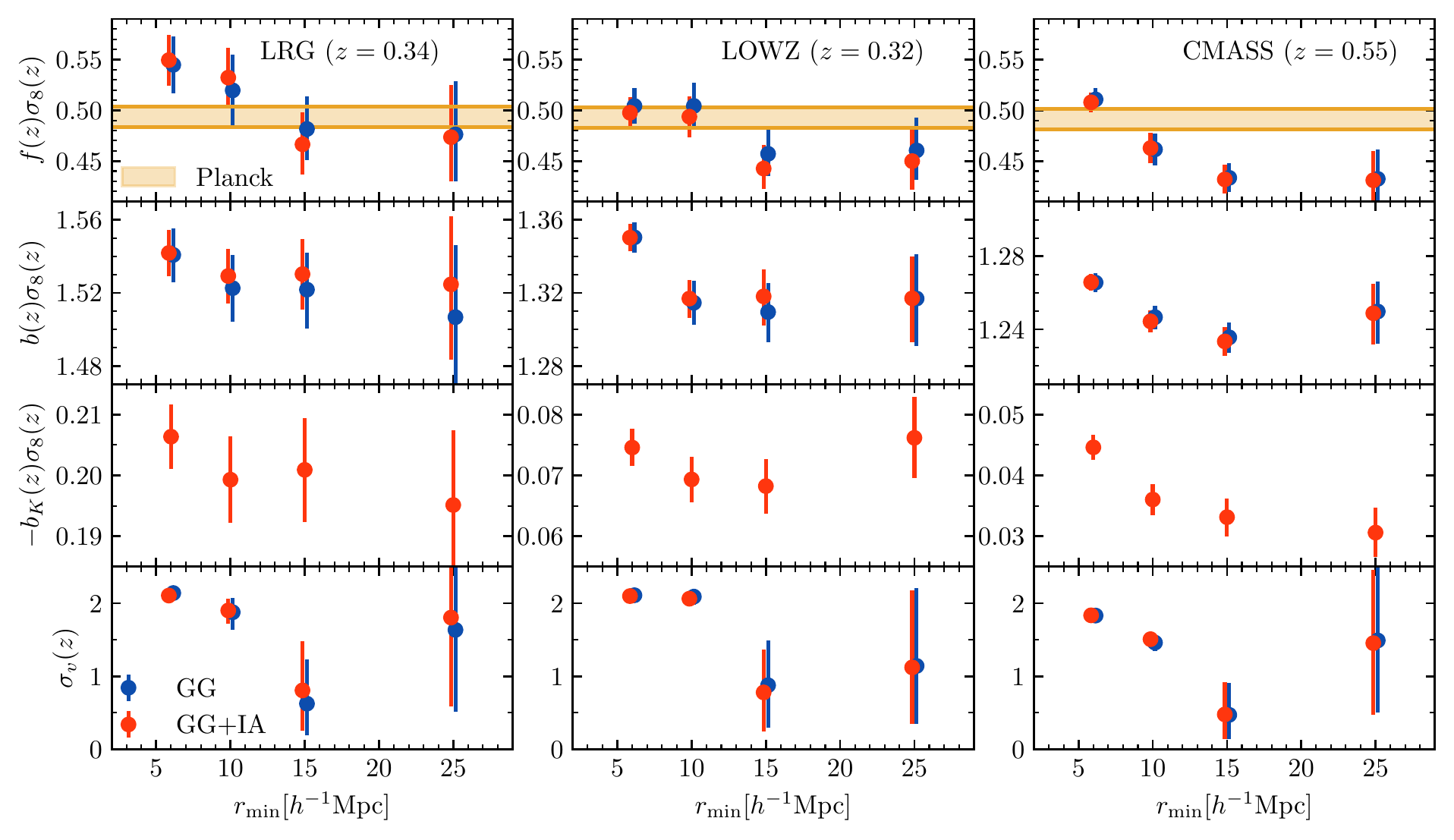} 
\caption{Constraints on model parameters as a function of the minimum separation, $r_{\rm min}$, obtained from clustering-only analysis and combined analysis of clustering and IA for LRG (left), LOWZ (middle) and CMASS (right) samples. We show the results for $f\sigma_8$, $b\sigma_8$, $b_K\sigma_8$, and $\sigma_v$ from the top to bottom rows. Theoretical prediction with 68\% C.~L. based on the Planck experiment is shown as the yellow regions in the top row.} \label{fig:scale_dependence} \end{center} \end{figure*}

In this appendix, we examine how cosmological constraints vary with the scales used in the likelihood analysis.  It is important because the growth rate constraint is prone to have scale-dependence due to various nonlinear effects \citep[e.g.,][]{Okumura:2011}.  The left column of Figure~\ref{fig:scale_dependence} shows the constraints on parameters for the LRG sample as a function of the minimum separation $r_{\rm min}$ after other three are marginalized over. The constraint on $f\sigma_8$ with the clustering-only analysis shows a strong scale dependence, with the same trend as the simulation result \citep{Okumura:2011}. 
The combined analysis of clustering and IA shows the same tendency.
Since the combined analysis with the scale cut of $r_{\rm min} = 10\mpcoh$ gives the best-fitting value of $f\sigma_8$ expected at the large scale limit ($25 < r < 100~[\mpcoh\ ]$), we present it as the main result of this paper.  The middle and right columns of Figure~\ref{fig:scale_dependence} show the scale dependence of parameter constraints obtained from the LOWZ and CMASS samples, respectively. The overall tendency of the constraints on $f\sigma_8$ is similar to that for the LRG sample. For consistency, we also adopt $r_{\rm min} = 10\mpcoh$ for the analysis of the LOWZ and CMASS samples.
However, as mentioned in Sec.~\ref{sec:result}, small error bars in the GG correlation of the CMASS sample result in the large $\chi^2$ value when we choose $r_{\rm min} = 10\mpcoh$ ($\chi^2/\nu=2.42$). If we adopt $r_{\rm min}=15\mpcoh$, the minimum $\chi^2$ is reduced to $\chi^2/\nu=1.68$. Accordingly, the best-fitting value of $f\sigma_8$ is shifted.

\section{Effect of PSF on parameter constraints}\label{sec:psf} As described in Sec.~\ref{sec:sdss}, the ellipticity of LRG is defined by the isophote of the light profile while that of LOWZ and CMASS galaxies is by the adaptive moment. \cite{Singh:2016} constructed the shape catalog for the LRG and LOWZ samples using a re-Gaussianization technique, which is based on the adaptive moment but involves additional steps to correct for non-Gaussianity of both the PSF and galaxy surface brightness profile \citep{Hirata:2003}. Utilizing it, \cite{Singh:2016} found that while the isophotal shape is not corrected for the PSF, the measured IA statistics are not so biased because the method uses the outer shape of the galaxies. Eventually, the uncorrected PSF affects only the amplitude of the measured IA statistics, not the shape, which has already been confirmed by our earlier work \citep{Okumura:2009}. Furthermore, \cite{Okumura:2009a} showed that the amplitude of IA, namely the shape bias $b_K$, determined by the GI and II correlations is fully consistent with each other. Therefore, while the constraint on $b_K$ can be different from the true value, that on the growth rate $f$ is not expected to be biased after $b_K$ is marginalized over. 
While the adaptive moment corrects for the PSF in the ellipticity, it results in a small bias \citep{Hirata:2003}. However, it is a constant bias, and thus it affects the amplitude of $b_K$, similarly to the isophotal shape definition but the effect is smaller. To be conservative, we exclude the II correlation at $r > 25 \mpcoh$ which is affected if we adopt the less accurate, de Vaucouleurs model fit\citep{Singh:2016}. Namely, the constraints from LOWZ and CMASS samples on $f\sigma_8$ with $r_{\rm min} = 25 \mpcoh$ in Fig.~\ref{fig:scale_dependence} do not use the data of the II correlation., Nevertheless, the constraints are almost equivalent to those with $r_{\rm min} = 15 \mpcoh$. It implies that the bias which arises from the uncorrected PSF is negligible for the shape definition of LOWZ and CMASS galaxies.

For all the three galaxy samples, constrained values of the model parameters do not change significantly by combining the IA statistics with the clustering statistics but shrink the error bars. It demonstrates that systematic effects associated with the shape measurement do not contribute to biases in the parameter constraints. More concrete discussion of uncorrected PSF effects on cosmological constraints requires the construction of shape catalogs in which the systematic effects are fully corrected for \citep{Hirata:2003,Singh:2016}. It will be investigated in future work.



\end{document}